# Spectral lens enables a minimalist framework for hyperspectral imaging

Zhou Zhou[1,5]†, Yiheng Zhang[2]†, Yingxin Xie[1]†, Tian Huang[1], Zile Li[1,3]*, Peng Chen[2]*, Yanqing Lu[2]*, Shaohua Yu[3]*, Shuang Zhang[6,7], Guoxing Zheng[1,3,4]*

[1]Electronic Information School, and School of Microelectronics, Wuhan University, Wuhan, 430072 China

[2]National Laboratory of Solid State Microstructures, Key Laboratory of Intelligent Optical Sensing and Manipulation, and College of Engineering and Applied Sciences, Nanjing University, Nanjing, 210093 China

[3]Peng Cheng Laboratory, Shenzhen, 518055 China

[4]Wuhan Institute of Quantum Technology, Wuhan, 430206 China

[5]NUS Graduate School, National University of Singapore, Singapore 119077, Singapore

[6]New Cornerstone Science Laboratory, Department of Physics, University of Hong Kong, Hong Kong, China

[7]Department of Electrical and Electronic Engineering, University of Hong Kong, Hong Kong, China

*Corresponding author. Email: lizile@whu.edu.cn; chenpeng@nju.edu.cn; yqlu@nju.edu.cn; yush@cae.cn; gxzheng@whu.edu.cn

†These authors contributed equally to this work

**Abstract:**

Conventional lens-based imaging techniques have long been limited to capturing only the intensity distribution of objects, resulting in the loss of other crucial dimensions such as spectral data. Here, we report a spectral lens that captures both spatial and spectral information, and further demonstrate a minimalist framework wherein hyperspectral imaging can be readily achieved by replacing lenses in standard cameras with our spectral lens. As a paradigm, we capitalize on planar liquid crystal optics to implement the proposed framework. Our experiments with various targets show that the resulting hyperspectral camera exhibits excellent performance in both spectral and spatial domains. With merits such as ultra-compactness and strong compatibility, our framework paves a practical pathway for advancing hyperspectral imaging apparatus toward miniaturization, with great potential for portable applications.

**Main Text:**

The diversity of ways in which humans acquire information directly influences our worldview and drives progress in technology and society. Optical imaging, the paramount method of information acquisition, has made significant advancements with the introduction of lenses, enabling the exploration into microscopic realms and distant universes. Yet conventional lens-based imaging techniques have been limited to capturing object intensities for centuries, neglecting other vital dimensions of information like spectral data, which is essential for understanding light-matter interactions in phenomena such as photon emission and molecular vibrations[1].

Spectrometers have been therefore developed as important instruments to capture spectral information, and they have played significant roles in diverse fields including material characterization[2], medical diagnosis[3], and remote sensing[4], etc. In order to acquire spatial and spectral information simultaneously, it is natural to consider a framework in which spectral filters and imaging components are cascaded together. However, the large footprints of the cascaded systems limit their applications in scenarios where miniaturization and portability are highly desired[5,6]. In recent years, researchers have explored various approaches to reduce the footprint of spectral detection systems[7–24]. However, these efforts have primarily focused on single-point spectral acquisition, such as the development of miniaturized spectrometers using quantum dots[7], nanowires[9–11], metasurfaces[13–16] and heterojunctions[19,20]. Although these miniaturized computational spectrometers can be arranged repeatedly on a chip[8,13,23] or scanned spatially[9,18–20] to capture spatial information (table S1), an additional lens is still required because the working units lack the phase manipulation capability for imaging, resulting in limited system integration. Moreover, achieving a balance between the performances in spectral and spatial domains remains a challenge in such frameworks. To overcome these obstacles and advance the miniaturization and integration of spectral imaging systems, innovative principles for light control and new system frameworks should be explored.

In this article, we propose a spectral lens which unifies the functions of optical imaging and computational spectrometry, thus enabling simultaneous spectral and spatial information acquisition of detected objects. By replacing the lenses in a standard camera with our spectral lens, hyperspectral imaging can be readily achieved. As a paradigm, we design the spectral lens using planar liquid crystal optics. The resulting hyperspectral camera exhibits excellent performance in both spectral and spatial domains, all within a compact footprint of 2.12 mm × 2.12 mm. Our experimental results demonstrate the high-quality acquisition of spectral images with 500 × 500 pixels, 29 spectral channels, and a spatial resolution of 31 μm. Notably, our framework achieves hyperspectral imaging with only a planar lens and a standard complementary metal-oxide-semiconductor (CMOS) sensor, contributing to a minimalist configuration with greatly reduced volume and weight. Our hyperspectral camera holds great potential for fulfilling the growing demands of portable applications, such as point-of-care diagnostics at home, advanced endoscopy and smart sensing in consumer electronics.

**Working principle of the spectral lens**

A lens that can achieve spectral detection while preserving its fundamental imaging functionality needs to fulfill two key criteria: (1) it should possess precise phase controls in a broad spectral range for high-quality imaging; (2) the focusing characteristics across different wavelengths should be as diverse as possible to better extract spectral information. Although these criteria can be satisfied by designing lenses with point spread functions (PSFs) that are spatially separated at different wavelengths[25,26], the spectral and spatial performances are mutually constrained in this approach (supplementary text section ST1).

It is well established that materials with innovative properties can enable devices with new functionalities, and the recent advancements in miniaturized spectrometers[7,9,14,20] have fully confirmed this wisdom. Liquid crystals (LCs), widely employed material in various modern technologies[27,28] due to their anisotropic and self-assemble characteristics, have invigorated renewed potential with the recent advancements in photoalignment technology, opening up possibilities for a series of novel photonic devices[29,30]. Here, we propose that the emerging liquid crystal planar optics offers a promising platform for developing the aforementioned lens. Spatially varying anisotropy in LC structures allows for precise phase manipulation through the spin-orbit interaction of light, known as the geometric phase[31–37]. Moreover, spectral modulation can be achieved by exploiting LCs' electrically tunable birefringence and dispersion[38–40]. These exceptional characteristics position LCs as a promising candidate to develop spectral lens that can meet the aforementioned criteria and address the challenges in current miniaturized spectral imaging systems.

Figure 1A illustrates the framework of spectral imaging with a liquid crystal spectral lens (LC-SLENS). By applying a voltage sequence on the LC-SLENS, different intensity image frames of the detected object can be captured with a CMOS sensor. The simultaneous spectral and spatial encoding with the LC-SLENS originates from the two degrees of freedom of the LC directors' orientation (Fig. 1B): the azimuth angle $\theta$ guided by the photoalignment agent (PAA), and the polar angle $\alpha$ tuned dynamically by the voltage applied across the LC layer. As shown in Fig. 1C, the spectral amplitude modulation remains unaffected by $\theta$ and only depends on $\alpha$[35]. For LCs with positive dielectric anisotropy, the LC directors tends to tilt along the applied electric field (parallel to the z-axis), so $\alpha$ becomes smaller as the voltage increases (supplementary text section ST2). On the other hand, the intended geometric phase modulation of an imaging lens is achieved by arranging the azimuth angle $\theta$, which remains decoupled from $\alpha$ (Fig. 1D). Such azimuth distribution is imprinted in the LC structures during photopatterning, and remains stable under different voltages. Consequently, the point spread function PSF$(x, y, \lambda)$ of the LC-SLENS remains unchanged while the spectral response $SR(V_i, \lambda)$ would vary according to the applied voltage. Hence, when a voltage $V_i$ is applied to the LC-SLENS, the captured intensity frame of an object with spatial-spectral information $O(x, y, \lambda)$ could be expressed by:

$$I_i(x, y) = \int_\Lambda SR(V_i, \lambda) \times [O(Mx, My, \lambda) \circledast \text{PSF}(x, y, \lambda)]\, d\lambda, \quad (1)$$

where Λ denotes the operating wavelength range, $M$ is the magnification of the LC-SLENS, and ⊛ represents the convolution operation in the *XOY* plane.

According to equation (1), the spectral and spatial information of the detected object is reflected in the images obtained with the LC-SLENS. We have therefore developed a two-step algorithm to recover the spectral datacube from the captured frames (Fig. 1E). Firstly, by discretizing equation (1) into a matrix form, we can reconstruct blurred spectral images (i.e., the term in the square bracket) point-by-point using convex optimization[8] (supplementary text section ST3). This reconstruction process relies on the pre-calibrated spectral response $SR(V_i, \lambda)$ and the image frames $I_i(x, y)$ obtained with the LC-SLENS. Subsequently, we employ Wiener filtering for deconvolution[41,42], leveraging the PSFs calibrated at each wavelength. By implementing these steps, the spectral information of all points on the object can be recovered.

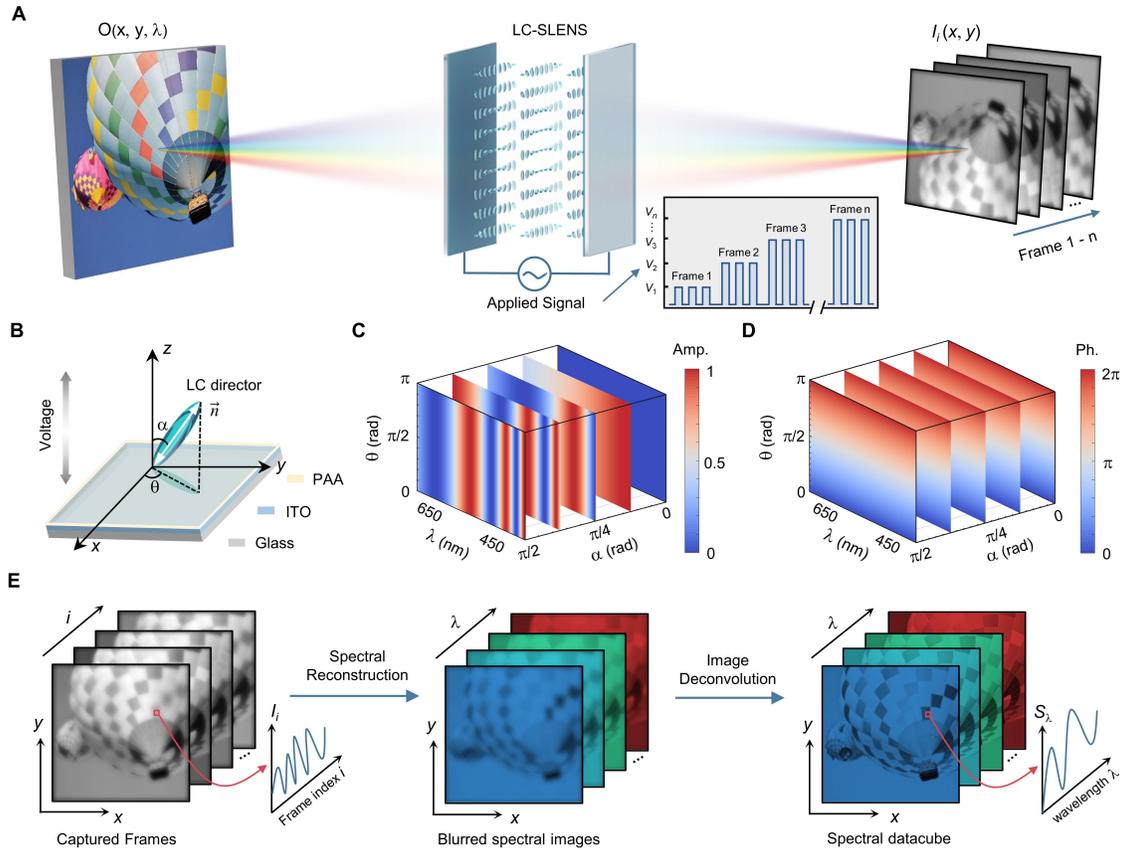

**Fig. 1. Concept and working principle of spectral imaging with a liquid crystal spectral lens (LC-SLENS).** (**A**) Data acquisition process of spectral imaging with the LC-SLENS. Varying frames of the detected object are captured by applying different voltages on the LC-SLENS. (**B**) Schematic illustration of an LC director with azimuth angle $\theta$ and polar angle $\alpha$. PAA: photoalignment agent. ITO: indium-tin-oxide glass. (**C** and **D**) Spectral amplitude (C) and phase (D) modulations by LCs versus $\theta$, $\alpha$ and wavelength $\lambda$. The color bars in (C) and (D) denote the normalized amplitude and relative phase of output light after passing through an LC layer (thickness: 6.3 μm), respectively. Five slices are shown to represent the modulation rule of LCs

under different *α* (different voltages). (**E**) Flowchart of a two-step algorithm to obtain the spectral datacube of the detected object. Blurred spectral images are firstly reconstructed from the captured frames by convex optimization, and deconvolution is then performed to attain clear spectral images.

For the experimental demonstration of spectral imaging with our proposed framework, we fabricated an LC device using a homemade setup (materials and methods). Figure 2A shows the polarized optical image of the fabricated LC device assembled with two indium-tin-oxide (ITO) glass substrates, and the external voltage across the LC layer is applied via the electrodes indicated by white dashed boxes. Each LC-SLENS on it has an aperture of 2.12 mm × 2.12 mm and a thickness of 6.3 μm. To balance the focusing performance of the LC-SLENS across the operating wavelength range, we employed a specially designed phase profile[25,43] and configured the azimuth distribution of the LC-SLENS as follows (supplementary text section ST4):

$$\theta(r) = k_r(f_0 - \sqrt{r^2 + f_0^2})/2, r^2 = x^2 + y^2 \qquad (2)$$

$$k_r = 2\pi/[\lambda_{min} + (\lambda_{max} - \lambda_{min})(\frac{r}{r_0})^2], \qquad (3)$$

where $f_0$ is the focal length, $\Lambda = [\lambda_{min}, \lambda_{max}]$ is the operating wavelength range and $r_0$ is the radius of the LC-SLENS. In the LC-SLENS design, we choose $f_0$ as 5 cm, and the operating wavelength range ($\Lambda$) spans from 550 nm to 700 nm. The resulting azimuth distribution of the designed LC-SLENS is shown in Fig. 2B.

To characterize the fabricated LC-SLENS, we first utilized a polarizing microscope with illumination from a halogen lamp filtered by bandpass filters (550 nm - 700 nm). Figure 2C demonstrates the microscopic image observed with a 5× objective, and the zoom-in-view in Fig. 2D was obtained with a 50× objective. These microscopic images show clear and high-contrast fringes, which agree well with our designed azimuth distribution (supplementary text section ST2). Next, we calibrated the spectral response of the LC-SLENS under different applied voltages using a commercial spectrometer (Thorlabs, CCS100). As shown in Fig. 2E, the LC-SLENS exhibits diverse spectral responses that vary with applied voltage, and this result conforms with our theoretical analysis (fig. S1). We note that the drop-off observed near 700 nm is attributed to the transmission spectral characteristics of the bandpass filters (fig. S5) used in calibration.

Subsequently, the PSFs at different wavelengths were measured using a pinhole and a super-continuum laser (YSL SC-pro). Figure 2F presents examples of the calibrated PSFs at 550 nm, 600 nm, 650 nm and 700 nm (refer to fig. S6 for PSFs across the operating wavelength ranges). Despite the presence of side lobes at longer wavelengths, the incident light focused by the LC-SLENS mostly concentrates around the center due to our specially designed phase profile, which guarantees a broad operation bandwidth for spectral imaging. Furthermore, the consistency of the LC-SLENS's PSFs under varying external voltages was experimentally verified (supplementary text section ST5). Analysis using correlation coefficients reveals that the distribution of PSFs remains constant, while the efficiency varies with the applied voltages (fig. S6 and S7). Notably, this variation matches with the calibrated spectral response shown in Fig. 2E, which validates that the data-acquisition process of the LC-SLENS can be well describe by equation (1).

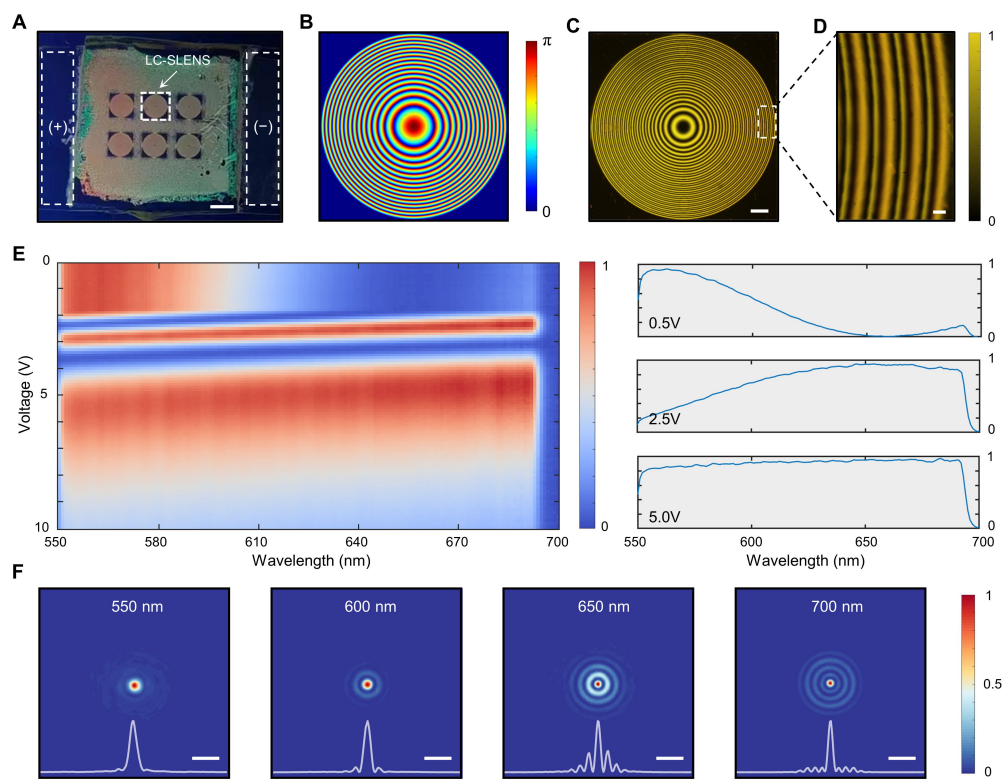

**Fig. 2. Characterization of the optical properties of the LC-SLENS.** (**A**) Optical image of the fabricated LC-SLENS (Scale bar: 2 mm). The electrodes for applying voltage are indicated by white dashed boxes. (**B**) Designed azimuth pattern (*θ*) of the LC-SLENS. (**C**) Cross-polarized micrograph of the LC-SLENS, obtained with a 5× objective. A halogen lamp filtered by bandpass filters (550 nm - 700 nm) is used for illumination. Scale bar: 200 μm. (**D**) Zoom-in view of the LC-SLENS's edges, obtained with a 50× objective. Scale bar: 20 μm. (**E**) Measured spectral responses of the LC-SLENS under different applied voltages. The spectra under 0.5 V, 2.5 V and 5.0 V are presented in the right panel to show the diverse voltage response of the LC-SLENS. (**F**) Measured PSFs of the LC-SLENS under 550 nm, 600 nm, 650 nm and 700 nm. The horizontal intensity profile along the center of each spot is denoted in the bottom. Scale bar: 100 μm.

**Spectral imaging with the proposed hyperspectral camera**

After calibrating the LC-SLENS's spectral response and PSFs, we combine it with a standard monochrome CMOS sensor (Omron, STC-MBS122BPOE) to form a hyperspectral camera (fig. S9). To assess the spectral accuracy of our proposed camera, we used a color board as the target, and a white LED source (~7500K) filtered by bandpass filters (550 nm - 700 nm) was employed for illumination. Figure 3A presents the raw images captured by our hyperspectral camera when different voltages are applied to the LC-SLENS (movie S1). Notably, the intensity contrast between different color blocks varies with the applied voltage due to the difference in their spectra. Using these image frames, we reconstructed the color board's spectral datacube and converted it to a color image (Fig. 3B). For reference, we measured the spectrum of each color block with a commercial spectrometer (Thorlabs, CCS100) and synthesized the corresponding color image (Fig. 3C). These two images show close agreement in hue and brightness for all blocks. Quantitatively, the spectral profiles of four representative color blocks captured by our camera (dotted lines) are plotted in Fig. 3D, which matches well with measurements from the commercial spectrometer (see fig. S10 for the data of other color blocks). These results validate that our camera can acquire spectral information accurately.

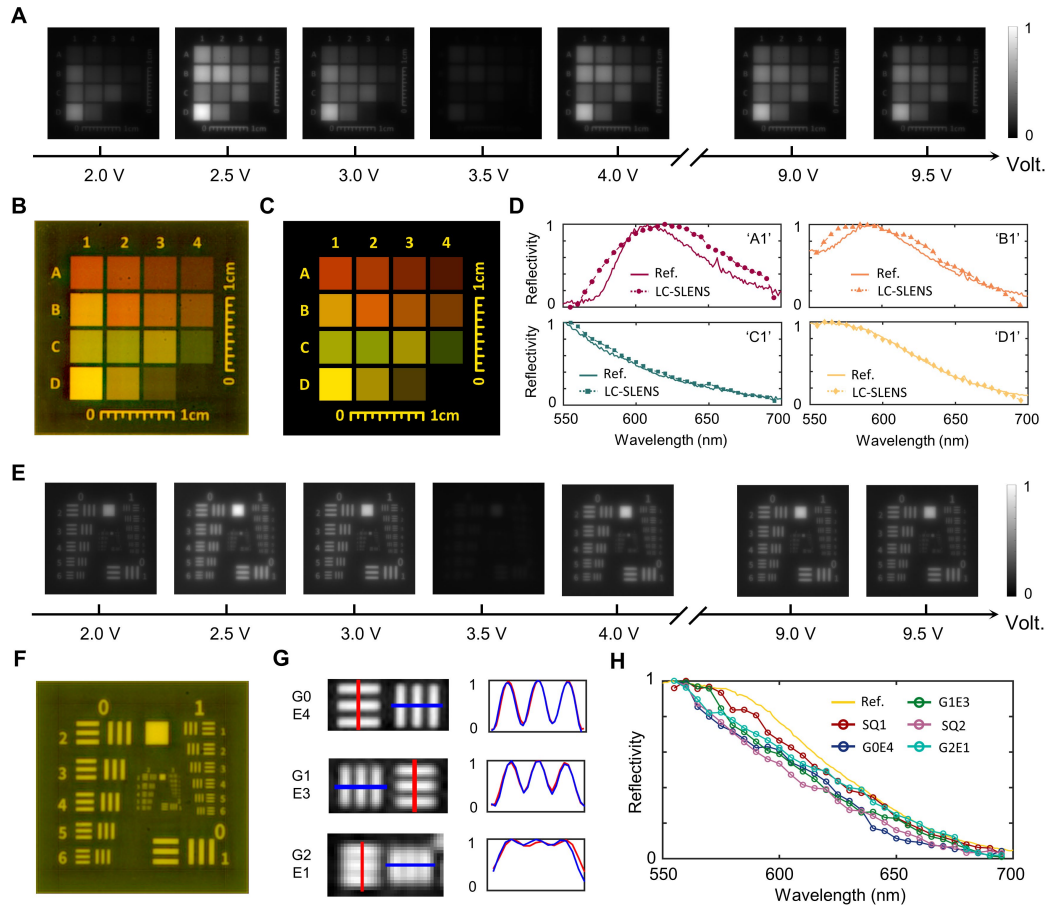

**Fig. 3. Spectral imaging tests of our proposed hyperspectral camera.** (**A**) Experimental images of a color board captured by applying different voltages on the LC-SLENS. (**B** and **C**) Synthesized color image of the color board by the spectra obtained with our hyperspectral camera (B) and a commercial spectrometer (C. Thorlabs, CCS100). (**D**) Spectral profiles of four blocks ('A1', 'B1', 'C1' and 'D1') on the color board. The dotted line and solid line represent the results obtained with our hyperspectral camera and the commercial spectrometer, respectively. (**E**) Experimental images of the USAF1951 resolution chart by applying different voltages on the LC-SLENS. (**F**) Synthesized color image of the resolution chart by the reconstructed spectral datacube. (**G**) Zoom-in grayscale image of three regions (Group 0 Element 4, Group 1 Element 3, Group 2 Element 1) on the resolution chart. The intensity distribution along the red and blue lines are demonstrated in the right panel. (**H**) Spectra at different positions on the resolution chart, with the spectrum on the big square (SQ1) measured using the commercial spectrometer as a reference. SQ2 represents the small square on the chart.

Moreover, the spatial resolution of our hyperspectral camera was tested with a USAF1951 resolution chart (under the same illumination condition as the color board experiment), and the corresponding raw data is presented in Fig. 3E. Due to the similarity in spectra across different regions of the chart, the intensity images obtained with the LC-SLENS exhibit overall variation as the voltage changes (movie S2). The reconstructed color image of the resolution chart (Fig. 3F) demonstrates nearly consistent colors for all line pairs, indicating the resemblance between their spectra. Zoom-in images of specific elements in Fig. 3G reveal that Group 2 Element 1 (4 lp/mm) can be successfully resolved, with a calculated resolution of 31 μm based on the image magnification ($M = -0.124$). Meanwhile, the reconstructed spectra at different positions on the resolution chart are in good accordance with the reference spectrum obtained using a commercial spectrometer (Fig. 3H). These results demonstrate our hyperspectral camera's ability to accurately capture spectral information of objects with fine spatial features.

We further demonstrate the versatility of our proposed hyperspectral camera by applying it in various scenarios. Firstly, the camera is employed to perform spectral imaging of a poster with the logo of Wuhan University. Figure 4A shows the poster image captured with a conventional RGB camera, which consists of an achromatic lens ($D = 25.4$ mm, $f = 75$ mm) and a color CMOS sensor (Omron, STC-MCS122BPOE). The synthesized color image from our hyperspectral camera (Fig. 4B) is fairly close to that obtained from the RGB camera, and the details in the logo can be revealed clearly. Although both these two cameras consist of a lens and a CMOS sensor, our hyperspectral camera is superior as it can provide extra spectral information (Fig. 4C) of the object. In addition, we captured the spectral images of a pattern representing Nanjing University displayed on a micro-LED screen (Unilumin, Umicro0.7). As can be seen from the pattern captured with an RGB camera (Fig. 4D), the green and red LEDs illuminate as "tree" and "NJU" respectively, while both LEDs light up at the pattern border (see fig. S11 for the pixel configuration of RGB LEDs). Figure 4E presents the color image obtained using our hyperspectral camera, which is consistent with the result from the RGB camera. Meanwhile, we can extract the images of the pattern at different wavelengths from the reconstructed spectral datacube. According to the LEDs' emission spectra (Fig. 4F) measured with a commercial spectrometer, the green LED emits stronger light than the red LED below 590 nm. Notably, this characteristic can be clearly verified in the normalized spectral images presented in Fig. 4G (see fig. S12 for more spectral images across the working band). These results showcase the capabilities and advantages of our hyperspectral camera in capturing spatial and spectral information in diverse scenarios.

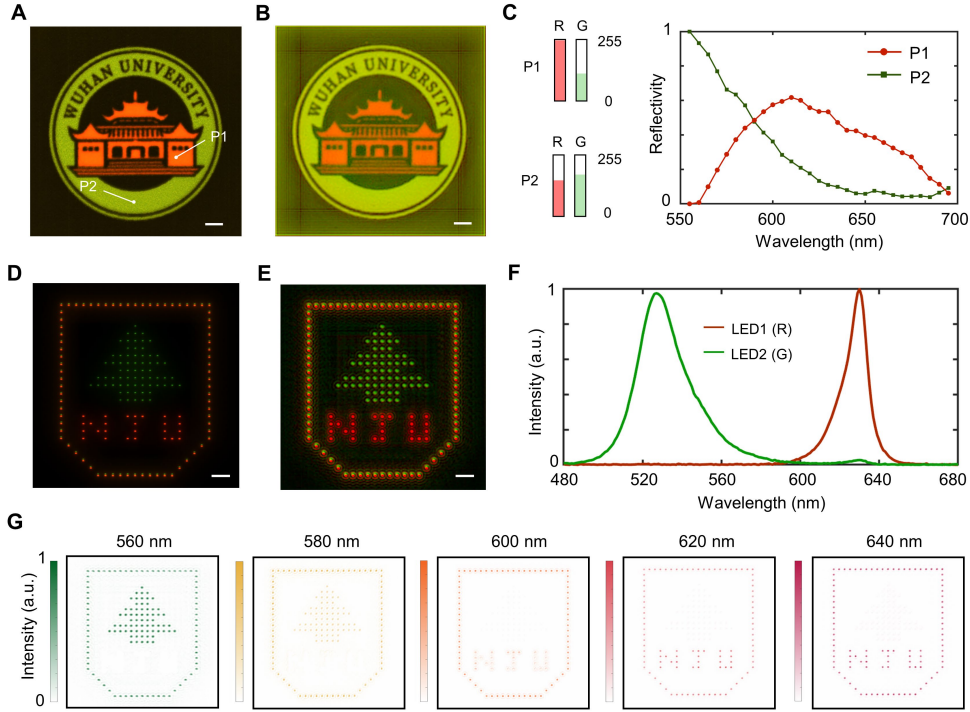

**Fig. 4. Demonstration of our hyperspectral camera's diverse applications.** (**A**) Color image of a poster (the logo of Wuhan University), captured with an RGB camera. (**B**) Synthesized color image of the poster, obtained using our hyperspectral camera. (**C**) The information of the poster acquired with the RGB camera (Left panel: intensity at P1 and P2 in R/G channel) and our hyperspectral camera (Right Panel: Spectra at P1 and P2). (**D**) Color image of a pattern (Nanjing University) displayed on a micro-LED screen (Unilumin, Umicro0.7), captured with an RGB camera. (**E**) Synthesized color image of the pattern, obtained using our hyperspectral camera. (**F**) Emission spectra of the green and red LEDs used in the micro-LED screen, measured with a commercial spectrometer. (**G**) Spectral images acquired by our hyperspectral camera at five distinctive wavelengths. Each image is normalized with its maximum intensity. Scale bar: 2 mm.

**Discussion and Conclusions**

In the 17th century, Sir Isaac Newton put forward the formula for lens imaging and carried out the color spectrum experiment, which were two important advancements in optics. Since then, lenses and spectrometers have been extensively studied as essential optical components for information acquisition. Here, the functionalities of these two distinct optical instruments are unified into one planar lens, enabling simultaneous spatial and spectral encoding. Therefore, the system configuration is significantly simplified and the overall integration is greatly enhanced. Notably, our proposed LC-SLENS exploits the azimuth pattern for imaging, while the polar orientation controlled by external voltages ensures spectral detection. The decoupled phase and spectral control allow us to overcome the constraints between spectral and spatial performance in previous frameworks. As a result, spectral imaging of objects with fine spatial structures can be achieved.

Moreover, in our framework, the information of all points within the spectral lens's field-of-view (FOV) is captured simultaneously, ensuring the acquisition of spectral images with large sizes. For spectral datacube reconstruction, we currently employ convex optimization and Wiener filtering. More advanced tools, such as deep learning[6,44], holds the potential to further enhance the spectral resolution and reconstruction accuracy, and the PSFs can also be optimized for better deconvolution, thereby improving the overall hyperspectral imaging performance. Lastly, we leveraged liquid crystals as a proof of concept to demonstrate the spectral lens, but the proposed framework can be further expanded upon through other novel materials (e.g., phase change materials and lithium niobate) with versatile light controls.

In conclusion, we have proposed a minimalist framework for hyperspectral imaging based on the concept of spectral lens. Within the platform of liquid crystal planar optics, we demonstrated a minimalist hyperspectral camera with excellent performance in acquiring both spectral and spatial information. The high efficiency, low cost, electrical tunability, ultrathin form factor, large-area manufacturing as well as the mature processing technology of the LC materials render our LC-SLENS promising for commercial applications. We envision broad application prospects of hyperspectral imaging with our proposed framework, particularly in areas where miniaturization, high integration and low system weight are highly desirable, such as smartphone and drone sensors, as well as portable healthcare devices. Furthermore, imaging systems with even more versatile functions can be created through synergy with other novel materials, such as incorporating with metasurfaces to achieve spectro-polarimetric imaging[16]. Overall, our proposed framework possesses strong compatibility and extendibility, which paves a practical way forward for approachable and efficient multi-dimensional information acquisition.

# Supplementary Materials for

**Spectral lens enables a minimalist framework for hyperspectral imaging**

**The PDF file includes:**

    Materials and Methods
    Supplementary Text
    Figs. S1 to S12
    Table S1
    References

**Other Supplementary Materials for this manuscript include the following:**

    Movies S1 and S2

**Materials and Methods**

MM1: Fabrication of the LC-SLENS

The fabrication of our LC device mainly involves two steps, i.e., preparation of the LC cell and ultraviolet (UV) photopatterning. Firstly, the ITO glass substrates are subjected to ultrasonic cleaning and UV-Ozone treatment. Next, a 0.3% solution of the sulphonic azo-dye SD1 (Dai-Nippon Ink and Chemicals, Japan) in dimethylformamide is spin-coated onto the substrates. After curing at 100 °C for 10 minutes, spacers with a thickness of 6 μm are dispersed over the SD1-coated substrate, followed by sealing the cell with epoxy glue using the other SD1-coated substrate. To transfer the desired azimuth pattern to the SD1 layer, we employ a DMD-based UV microlithography system comprising a light source, components for dynamic pattern generation and focusing, and a monitor. Specifically, a UV beam carrying the designed pattern is reflected onto the DMD (Discovery 3000, Texas Instruments), and it is then focused by a tunable lens, polarized by a motorized polarizer, and finally projected onto the LC cell. Upon injecting the E7 LCs, the patterned SD1 can effectively guide the orientation of the LC directors through intermolecular interactions, resulting in a fully assembled LC cell with the desired azimuth orientations.

MM2: Spectral response and PSF calibration

To calibrate the spectral responses (cross-polarization efficiency) of the LC-SLENS (Fig. S4a), we utilized a broadband light source (Thorlabs, MNWHL4) and a commercial spectrometer (Thorlabs, CCS100). Prior to illuminating the LC-SLENS, we employed a bandpass filter (BPF) with a passband of 550 nm to 700 nm (LBTEK, MEFH10-550LP and MEFH10-700SP). An iris was placed in front of the LC-SLENS to minimize the influence of stray light. In addition, a left-handed circular polarizer (CP1) and a right-handed circular analyzer (CP2) were utilized to eliminate the unwanted co-polarized light. We note that the spectral response of the BPF, CP1, LC-SLENS, and CP2 is measured as a whole during the calibration. The light passing through these components was collected by an aspheric condenser lens (Thorlabs, ACL4532U) for spectrum measurement. By applying square wave signals with a frequency of 1 kHz and varying peak-to-peak amplitude ($V_i$) generated by a signal generator (Tektronix, AFG31052) to the LC-SLENS, we obtained a series of spectra $S(V_i, \lambda)$ at different applied voltages. After removing the BPF, CP1, CP2, and LC-SLENS in the setup, we measured the spectrum of the incident light $S_0(\lambda)$.

Consequently, the spectral response of the system was calculated as $SR(V_i, \lambda) = S(V_i, \lambda)/S_0(\lambda)$.

For the calibration of the LC-SLENS's PSFs (Fig. S4b), we positioned a pinhole with a diameter of 20 μm (Thorlabs, P20K) in front of the LC-SLENS at a distance of 50 cm, and the CMOS sensor was placed 5.5 cm away from the LC-SLENS. To obtain the PSFs at different wavelengths, we illuminated the pinhole with a super-continuum laser (YSL SC-pro) and collected the corresponding images. A voltage of 5 V was applied to the LC-SLENS to ensure high conversion efficiency across the entire working band. Since the change in PSFs over the wavelength was insignificant benefitting from the designed phase profile, we adjusted the laser wavelength with a step of 10 nm and captured 16 PSF images corresponding to the operational wavelength range of 550 nm to 700 nm.

MM3: Converting spectral datacube to color image

To visualize the 3D spectral datacube of objects, we can convert it to a 2D color image. For each point on the object, its spectrum $S(\lambda)$ is used to calculate the CIE 1931 XYZ values (2° observer) using the equation $T_i = \sum_\Lambda S(\lambda) \bar{t}_i(\lambda) \Delta\lambda$ ($i$ = 1, 2, 3), where $\{\bar{t}_i(\lambda)\}$ denotes the CIE 1931 color matching functions, and $\Delta\lambda$ is the wavelength gap in calculation. These XYZ values are then converted into the sRGB color space using a transformation matrix. By repeating this process for all pixels in the spectral datacube, the corresponding color image can be synthesized.

**Supplementary Text**

ST1. Comparison between different schemes of miniaturized spectral imaging

Spectral imaging instruments consist of two essential parts: the optical system (front end) and the detection sensor (back end). Existing approaches for spectral imaging can be categorized based on whether the key working component belongs to the front or back end of the system. In this section, we provide a brief review of previous methods employed in miniaturized spectral imaging systems and compare them to our proposed spectral lens framework (Table S1).

- Miniaturized spectral imaging with front-end components

Traditional spectral imaging frameworks commonly employ dispersive elements/color filters and lenses in the optical system. One strategy for system miniaturization is to leverage the ultra-compact nature of metasurfaces and develop filters and lenses based on this novel hardware platform. For instance, by cascading a Fabry-Pérot cavity array and a metalens array[17], the weight and size of the spectral imaging system can be significantly reduced. This configuration is essentially an analog of conventional wavelength-scanning spectral imaging systems, and the device footprint and spectral performance are mutually constrained. Another approach is designing lenses with distinct focusing characteristics at different wavelengths, such as lenses with PSFs exhibiting different positions in space for varying wavelengths[25] or utilizing the transverse dispersion of off-axis lenses[26]. While these approaches require only one lens for the optical system, both spectral and spatial encoding are achieved through the PSF design. This coupling between the two encodings would lead to tradeoffs among spectral resolution, spatial resolution, and the size of spectral images.

- Miniaturized spectral imaging with back-end components

To reduce the footprint of spectral imaging systems in the back end, numerous miniaturized spectrometers have been reported by capitalizing on nanophotonic platforms, including quantum dots[7], photonic crystals[8], nanowires[9], black phosphorus[12], metasurfaces[13-16], perovskite[18], heterojunctions[19,20], and Fabry-Pérot cavity[23]. The general idea behind these approaches is to construct nanophotonic filters with distinct spectral responses by carefully engineering the unit cells. Since these unit cells are generally on the order of one wavelength, the resulting supercell for spectral detection is small enough to be arranged repetitively on a chip to realize spectral imaging. However, in such frameworks, the size of one detection pixel increases if higher spectral reconstruction performance is desired, leading to a degradation of spatial resolution. Additionally, the size of spectral images is directly determined by the repeated number of supercell detection units, which incurs higher costs for larger arrays. Furthermore, spectral imaging with these back-end components still requires a bulky lens, which impedes overall system miniaturization and integration.

In our proposed framework, the optical systems involve only one single lens, and a standard CMOS sensor is used for detection. Both the components used in the front and back ends are minimized, resulting in a minimalist system configuration and enhanced integration. Notably, spatial encoding in our spectral lens framework is achieved through liquid crystals' phase control, while spectral encoding is realized by their electrically tunable spectral manipulation. These two decoupled controls allow us to overcome the constraints between spectral and spatial performance. Furthermore, our spectral lens framework enables the capture of information from all points within the lens's field-of-view (FOV), ensuring the acquisition of spectral images with large sizes in a single measurement period. Consequently, our spectral lens framework enables spectral imaging with high spectral and spatial performance.

ST2. Light modulation with LCs

Considering the light propagating along the *z*-axis, the nematic liquid crystal (LC) is characterized by the director with azimuth angle $\theta$ and polar angle $\alpha$, as illustrated in Fig. 1B. This LC device can be regarded as a waveplate with optical axis oriented at $\theta$, and its phase retardation $\eta$ depends on $\alpha$:

$$\eta = \frac{2\pi(n_{eff} - n_o)d}{\lambda} \tag{S1}$$

where $n_{eff}$ and $n_o$ denote the refractive indices associated with the extraordinary and ordinary light wave respectively, $\lambda$ is the wavelength of incident light, and $d$ is the thickness of the LC layer. Here, $n_{eff}$ changes along with the polar angle $\alpha$, and it can be calculated with the following equation:

$$n_{eff} = \frac{n_o n_e}{\sqrt{(n_o \sin\alpha)^2 + (n_e \cos\alpha)^2}} \tag{S2}$$

We can see that when the LC director is in the *XOY* plane (Fig. 1B), i.e., $\alpha = \pi/2$, the anisotropy ($\Delta n = n_e - n_o$) is the strongest. Conversely, when the LC director is parallel to the *z*-axis, the material becomes equivalent to isotropic media. By applying a voltage along the *z*-axis, the LC director tends to align parallel to the electric field, and $\alpha$ would decrease from $\pi/2$ to 0 as the voltage increases. This characteristic enables the LC device to function as a waveplate with electrically tunable retardation[45]. The light modulation properties of the LC material are determined by its retardation controlled by $\alpha$, and the azimuth angle $\theta$, in the form of Jones matrix:

$$J_{LC} = \begin{bmatrix} \cos\theta & -\sin\theta \\ \sin\theta & \cos\theta \end{bmatrix} \begin{bmatrix} 1 & 0 \\ 0 & e^{i\eta(\lambda,\alpha)} \end{bmatrix} \begin{bmatrix} \cos\theta & \sin\theta \\ -\sin\theta & \cos\theta \end{bmatrix} \tag{S3}$$

With equation S3, we can derive the simultaneous spectral amplitude and phase modulation by LCs. For the incidence of left-handed circularly polarized (LCP) light $|L\rangle = [1, i]^T$, the modulation of right-handed circularly polarized (RCP) component in the output light can be calculated by:

$$\tilde{A} = \langle R | J_{LC} | L \rangle = \frac{1 - e^{i\eta(\lambda,\alpha)}}{2} e^{i2\theta} \tag{S4}$$

Notably, the RCP component is attached with an extra phase term $\exp(i2\theta)$, known as the geometric phase or Pancharatnam-Berry phase. Such phase modulation depends purely on the LC directors' azimuth angle $\theta$, regardless of incident wavelength. Meanwhile, the amplitude of this component depends on the wavelength of incident light and the polar angle of LC directors, resulting in spectral modulation. The azimuth angle of the LC directors can be individually controlled through the photopatterning process, while the variation in the polar angle is uniform across all pixels due to the identical applied voltage. Consequently, we can establish the following equations to describe the phase modulation and spectral modulation in the LC device:

$$\varphi(x,y) = 2\theta(x,y) \tag{S5}$$

$$SR(\lambda,\alpha) = \left|\frac{1-e^{i\eta(\lambda,\alpha)}}{2}\right|^2 = \sin^2\left\{\frac{\pi d}{\lambda}\frac{n_o[n_e - \sqrt{(n_o\sin\alpha)^2 + (n_e\cos\alpha)^2}]}{\sqrt{(n_o\sin\alpha)^2 + (n_e\cos\alpha)^2}}\right\} \tag{S6}$$

where (*x*, *y*) denotes the position of different working units. From equations S5 and S6, we can see that the phase modulation only depends on the azimuth pattern of the LC device. And the spectral modulation is fully decoupled from the phase control, which is only determined by the polar angle (applied voltage). To demonstrate the spectral modulation with LCs (E7, 25°C, thickness: 6.3 *μm*), we simulated the change of spectral modulation versus *α* according to equation S6, and the result is shown in Fig. S1.

In addition, intensity modulation can be achieved when the LC device is placed in an orthogonal polarization setup. If incident light with polarization state $|E_{in}\rangle = [\cos\beta, \sin\beta]^T$ passes through the LC device, and a bulk-optic analyzer with polarization direction orthogonal to that of the incident light, the output light intensity can be calculated by:

$$I = |\langle P|J(\theta)|E_{in}\rangle|^2 = SR(\lambda,\alpha)\sin^2(2\theta - 2\beta) \tag{S7}$$

where $\langle P| = [-\sin\beta, \cos\beta]$ denotes the polarization filtering of the analyzer.

From this equation, we can see that the intensity can be controlled point by point through configuring the azimuth angle of local LC directors. Therefore, an intensity image can be observed by placing a photopatterned LC device in a polarizing microscope. The normalized intensity distribution of the observed image keeps the same regardless of the wavelength of incident light. However, for broadband illumination with different light sources, the observed patterned LC device would exhibit varying colors due to the spectral response term in equation S7.

Benefitting from this intensity modulation characteristic, we can evaluate whether the photopatterned azimuth distribution is accurate by the LC device's cross-polarized micrograph. According to the designed azimuth pattern (Fig. S2A), an intensity image that would be observed in orthogonal optical setup can be calculated with equation S7, as shown in Fig. S2B (*β* = 0). We can see that this result is very close to the cross-polarized micrograph of the fabricated LC device (Fig. S2C), both of which exhibits 32 circular fringes.

ST3. Spectral datacube reconstruction algorithms

The image frames obtained with the LC-SLENS contain the information of the detected object's spectral datacube $O(x, y, \lambda)$. Therefore, we need to design a reconstruction algorithm to solve the inverse problem of the following equation:

$$I_i(x, y) = \int_\Lambda LC(V_i, \lambda) \times P(\lambda) \times D(\lambda) \times [O(Mx, My, \lambda) \circledast PSF(x, y, \lambda)] \, d\lambda \tag{S8}$$

where $LC(V_i, \lambda)$, $P(\lambda)$ and $D(\lambda)$ denote the spectral response of the LC-SLENS, the optical path and the CMOS sensor respectively. These responses can all be pre-calibrated and they form the system's spectral response $SR(V_i, \lambda) = LC(V_i, \lambda) \times P(\lambda) \times D(\lambda)$. By regarding the term in the square bracket as $O_b(x, y, \lambda)$, the above problem can be divided into two parts.

(1) Spectral reconstruction

The first step is to recover $O_b(x, y, \lambda)$ from the image frames $\{I_i(x, y), i = 1, 2, 3...\}$ and the spectral response $SR(V_i, \lambda)$, and their relationship can be expressed as:

$$I_i(x, y) = \int_\Lambda SR(V_i, \lambda) \times O_b(x, y, \lambda) \, d\lambda \tag{S9}$$

In the recovery process, the spectrum of each point on the detected object can be discretized into an $Q \times 1$ vector $\boldsymbol{O_b} = [O_b(\lambda_1), O_b(\lambda_2), ..., O_b(\lambda_Q)]^T$, and the spectral response forms an $P \times Q$ measurement matrix $\boldsymbol{SR}$, where $P$ is the number of captured intensity frames under different voltages. This leads to $P$ linear equations with $Q$ unknowns for each point in the captured frames:

$$\boldsymbol{I} = \boldsymbol{SR} \cdot \boldsymbol{O_b} \tag{S10}$$

Due to the experimental noise and smooth nature of the measured spectrum, we use $l_2$ norm and a regularization term with a weight $\alpha$ to estimate $\boldsymbol{O_b}$:

$$\min_{\boldsymbol{O_b}} \|\boldsymbol{I} - \boldsymbol{SR} \cdot \boldsymbol{O_b}\|_2^2 + \alpha \|\boldsymbol{O_b}\|_2^2, \text{ subject to } \boldsymbol{O_{bi}} > 0 \tag{S11}$$

By solving Equation S11 for each point on the captured frame[46], we can get the blurred spectral images $O_b(x, y, \lambda)$.

(2) Image Deconvolution

This step aims to obtain clear spectral images $\hat{O}(x, y, \lambda)$ by solving the inverse problem of the equation shown below:

$$\hat{O}(x, y, \lambda) = O_b(x, y, \lambda) \circledast PSF(x, y, \lambda) \tag{S12}$$

By using $O_b(x, y, \lambda)$ recovered in the last step and the point spread functions $PSF(x, y, \lambda)$, we can utilize Wiener filtering for deconvolution. For each wavelength $\lambda_0$, equation S13 can be used to deblur the image.

$$\hat{O}(x,y) = \mathcal{F}^{-1}\left\{\frac{H^*(u,v)}{|H(u,v)|^2 + 1/S}G(u,v)\right\} \qquad (S13)$$

where $\mathcal{F}^{-1}$ denotes the inverse Fourier transform and $*$ is the conjugate operator. In this equation, $H(u, v)=\mathcal{F}\{PSF(x, y, \lambda_0)\}$, $G(u, v)=\mathcal{F}\{O_b(x, y, \lambda_0)\}$, and $S$ is a constant estimating the signal to noise ratio. By applying Wiener filtering, the spatial resolution and the overall quality of the reconstructed spectral images can be enhanced.

## ST4. PSF design and MTF analysis

Our LC-SLENS works with geometric phase and its phase distribution remains unchanged with wavelength variation. Without special phase design, the lens can only focus light at a particular wavelength. Here we perform annular partition of the LC-SLENS to enable its functionality across the operating spectral range $\Lambda = [\lambda_{min}, \lambda_{max}]$. Specifically, the phase distribution of the LC-SLENS is as follows:

$$\varphi(r) = \frac{2\pi}{\lambda(r)}(f - \sqrt{r^2 + f^2}) \tag{S14}$$

where $r$ denotes the radial position on the LC-SLENS and $f$ is the focus length. To balance the focusing performance across the working bandwidth, each ring element (Fig. S3) is designed to have the same area. By setting the wavelength at the center and the outer edge as $\lambda_{min}$ and $\lambda_{max}$ respectively, the wavelength distribution $\lambda(r)$ can be derived as:

$$\lambda(r) = \lambda_{min} + (\lambda_{max} - \lambda_{min})\left(\frac{r}{r_0}\right)^2 \tag{S15}$$

where $r_0$ is the radius of the LC-SLENS.

With this specific PSF design, there is always an annular ring on the LC-SELNS that can focus the light within the operation range. Although the other areas would contribute to side lobes in the focal spot, the image blur can be solved through Winner filtering. Compared to traditional design in which $\lambda(r)$ is a constant, although the focusing performance at each wavelength is compromised in this PSF design, the performance across the operating bandwidth is more balanced, allowing for broadband spectral imaging.

Figure S6A demonstrates the measured PSFs from 550 nm to 700 nm, with a wavelength step of 10 nm. Several features can be observed in these PSF images. Firstly, the PSFs are centrosymmetric, so spatial resolutions along different directions should be similar, as experimentally demonstrated in Fig. 3G. Secondly, the PSFs show smooth change with wavelength variation, and the energy in the focal spot is mostly concentrated in the central region. These characteristics align with our PSF design.

To quantitatively evaluate the imaging performance of our LC-SLENS. We calculated the MTF curve based on the PSFs shown in Fig. S6A. Considering the numerical aperture (NA) of the LC-SLENS, we calculated the Rayleigh diffraction limit at different wavelengths and denote it with red dashed lines in the MTF curves. Additionally, the spatial frequencies corresponding to four times the Rayleigh diffraction limit are marked with blue lines. We can see that the nearly all MTF curves keep non-cut-off in the region denoted with blue line. This indicates that with our PSF design, the LC-SLENS can maintain satisfactory performance in a broad spectral range although the resolution at each wavelength is slightly compromised.

ST5. Variation of the PSF versus applied voltage

In the PSF measurement, the object is a point source with a specific wavelength $\lambda_0$, and its intensity distribution can be represented by a Dirac delta function $\delta(x, y, \lambda_0)$. Therefore, according to equation S8, the captured image can be described by:

$$I_i(x, y) = SR(V_i, \lambda_0) \times PSF(x, y, \lambda_0) \tag{S16}$$

From this equation, we can see that the image captured under different applied voltages should exhibit the same distribution, and the relative intensity variation between them follows the spectral response. To experimentally demonstrate this property, we applied a voltage sequence {0.5 V : 0.5V : 1.5 V, 1.6 V : 0.1 V : 5 V, 5.5 V : 0.5V : 10V} to the LC-SLENS and several PSF images are captured with the same exposure time. As an example, the PSF images at 630 nm are demonstrated in Fig. S7, in which we can observe the variation in the maximum intensity and the nearly constant intensity distribution.

According to equation S16, the spectral response at $\lambda_0$ can be obtained by summing the grayscales of all pixels on the captured PSF images, i.e., $\sum_x \sum_y I_i(x, y)$. Figure S8A shows the comparison between this result (blue asterisk line) and that measured using a broadband light source and a spectrometer (red line). These two curves agree well with each other, which demonstrates the effectiveness of equation S16 for describing the PSF measurement process. Meanwhile, we calculated the Pearson correlation coefficient $CC_{ij}$ of the PSF images obtained under different voltages, and obtained the map shown in Fig. S8B. Except for the two voltages with the lowest spectral response (and thus drastically affected by the noise), all the correlation coefficients are higher than 0.9. This confirms that the PSF distribution keeps the same regardless of the applied voltage, and the imaging function of the LC-SLENS is fully decoupled from its spectral tunability.

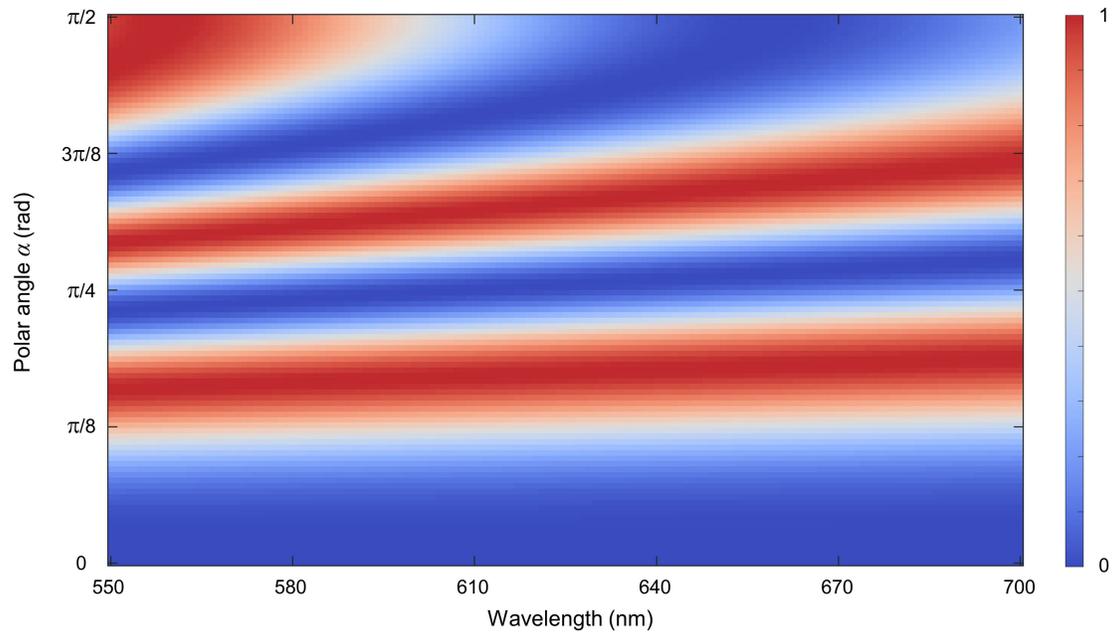

**Fig. S1**. **Spectral modulation of liquid crystals versus polar angle.** The color bar denotes the normalized amplitude of output light after passing through an LC layer (thickness: 6.3 μm). The birefringence index of E7 liquid crystals[47] at 25 °C is used for simulation.

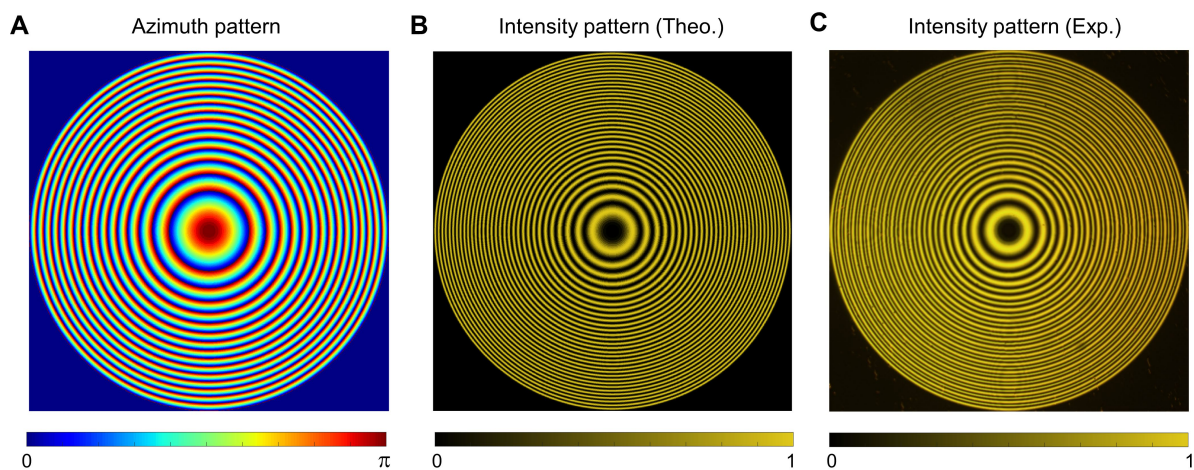

**Fig. S2**. **Intensity modulation with azimuth angle of liquid crystals.** (**A**) Designed azimuth pattern (36-step quantified) of the LC device. (**B**) Theoretical intensity pattern of the LC device when it is placed in an orthogonal optical setup ($\beta = 0$). (**C**) Experimentally observed intensity pattern of the LC device using a polarizing microscope.

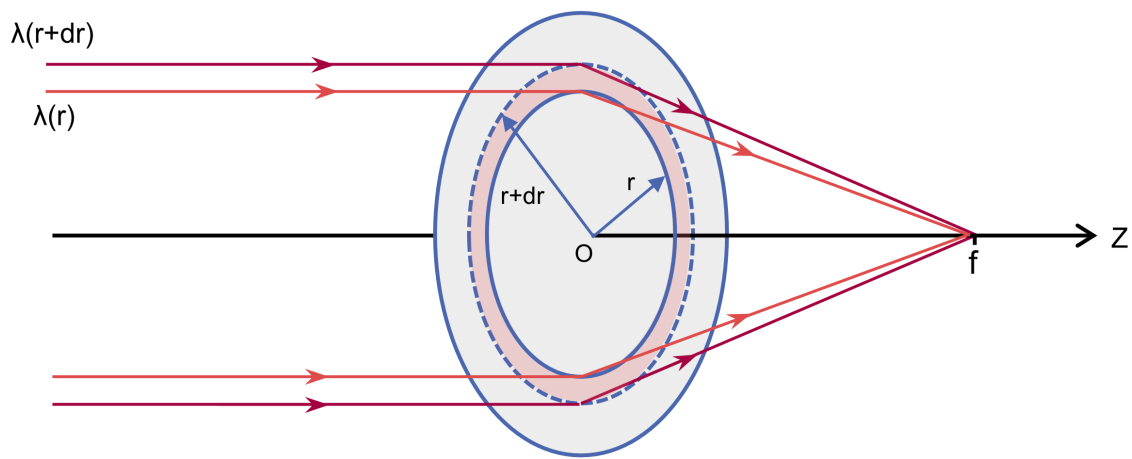

**Fig. S3. PSF design of the LC-SLENS.** The lens is divided into several annular ring elements with equal area and each region is responsible for focusing light with a particular wavelength.

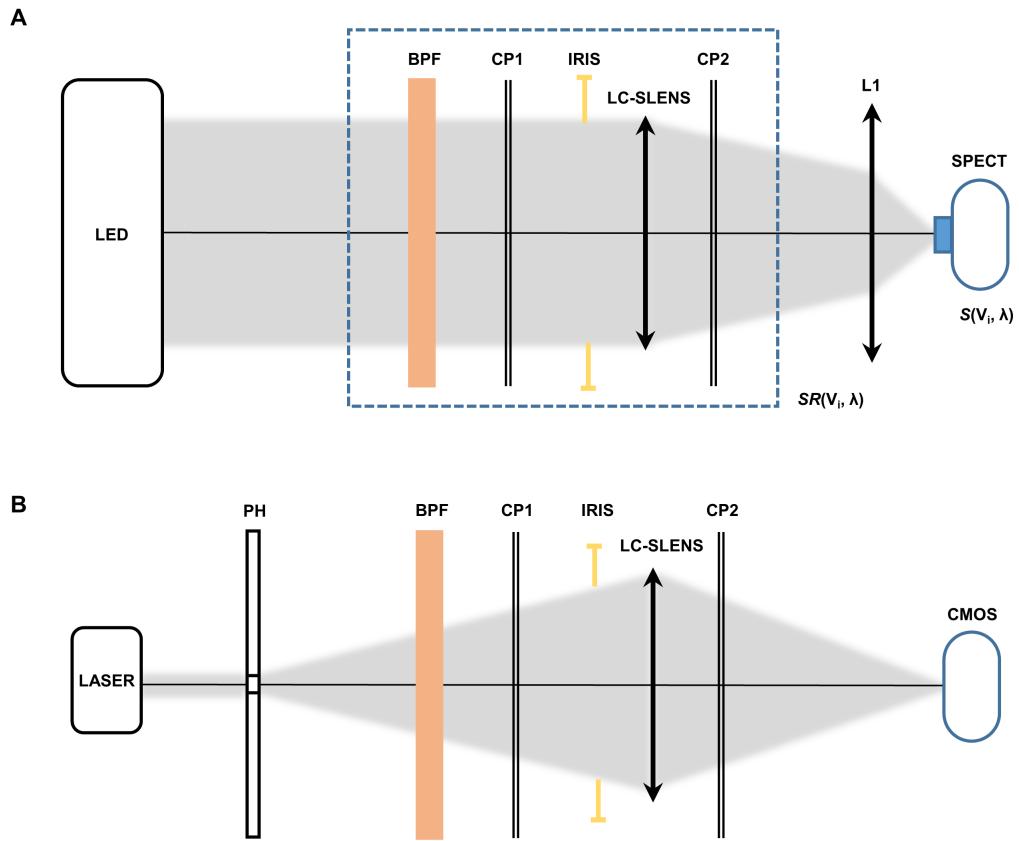

**Fig. S4**. **Schematic illustration of the optical setup for calibrating (A) spectral response and (B) PSFs.** BPF: bandpass filter, CP1/CP2: circular polarizer, SPECT: spectrometer, PH: pinhole.

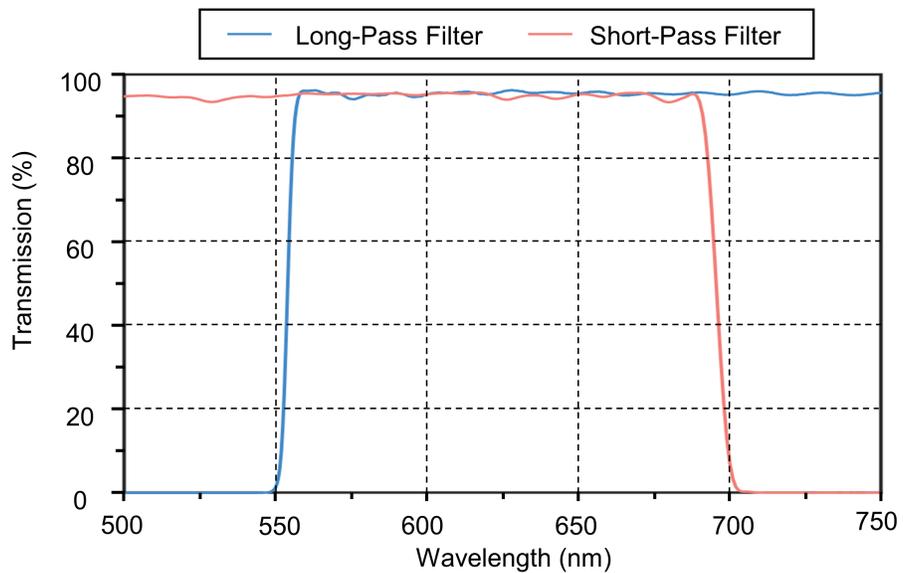

**Fig. S5**. **Transmission spectra of the bandpass filters (BPF) used in the system.** The BPF is composed of a long-pass filter and a short-pass filter, and their transmission spectra are denoted by blue and red lines, respectively.

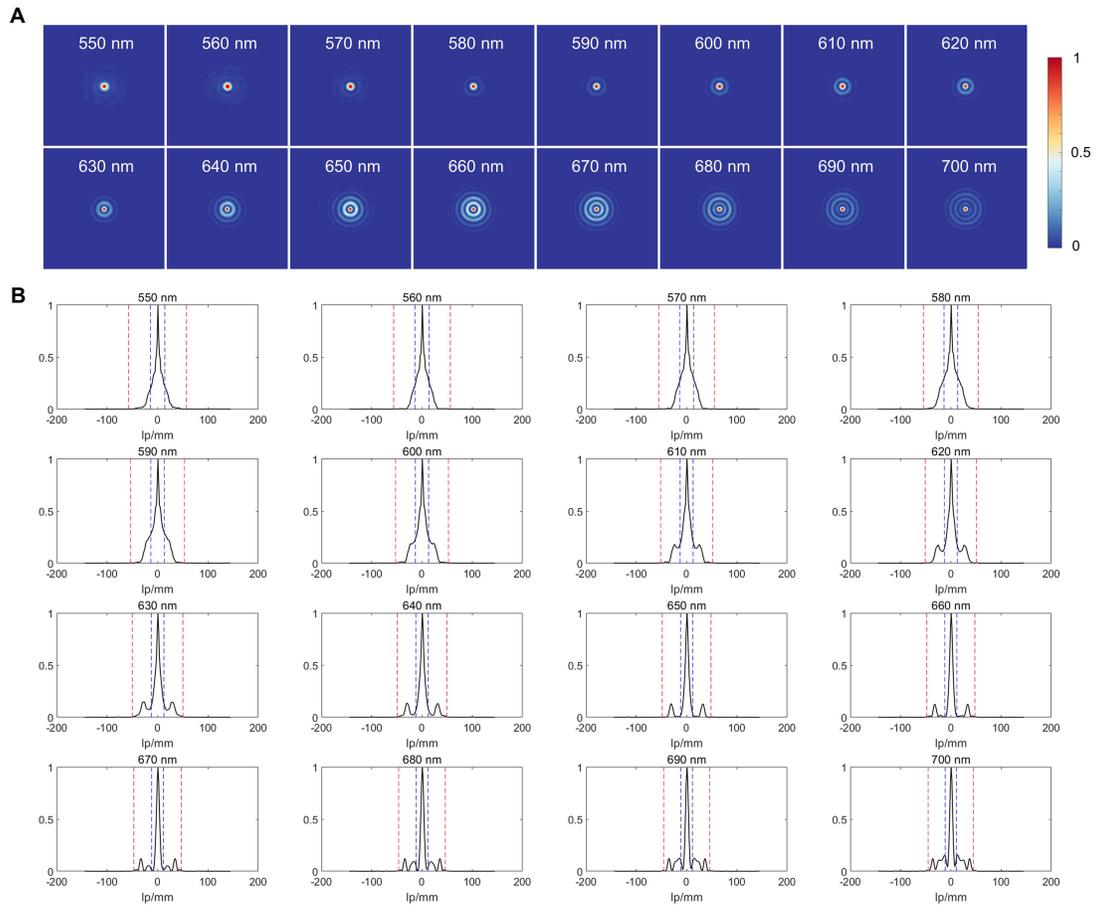

**Fig. S6**. **Focusing performance of the LC-SLENS across the operating range.** (**A**) PSFs at different wavelengths. (**B**) MTF curves at different wavelengths. The dashed red line and blue line denote the spatial frequencies corresponding to the diffraction limit and four times the diffraction limit, respectively.

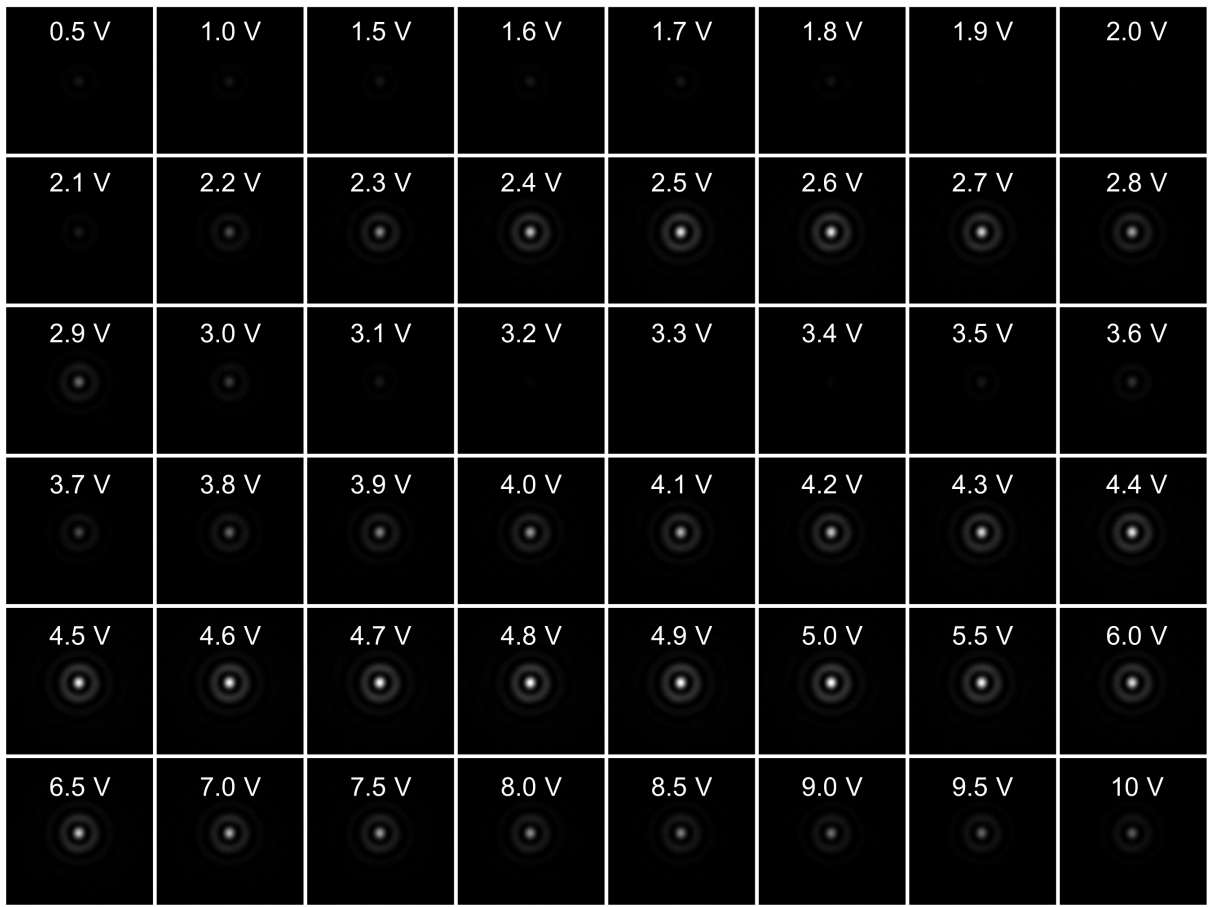

**Fig. S7**. **Variation of the PSF image versus applied voltage.** The exposure time of the CMOS sensor keeps the same during the capture of these images. The wavelength of incident light is 630 nm.

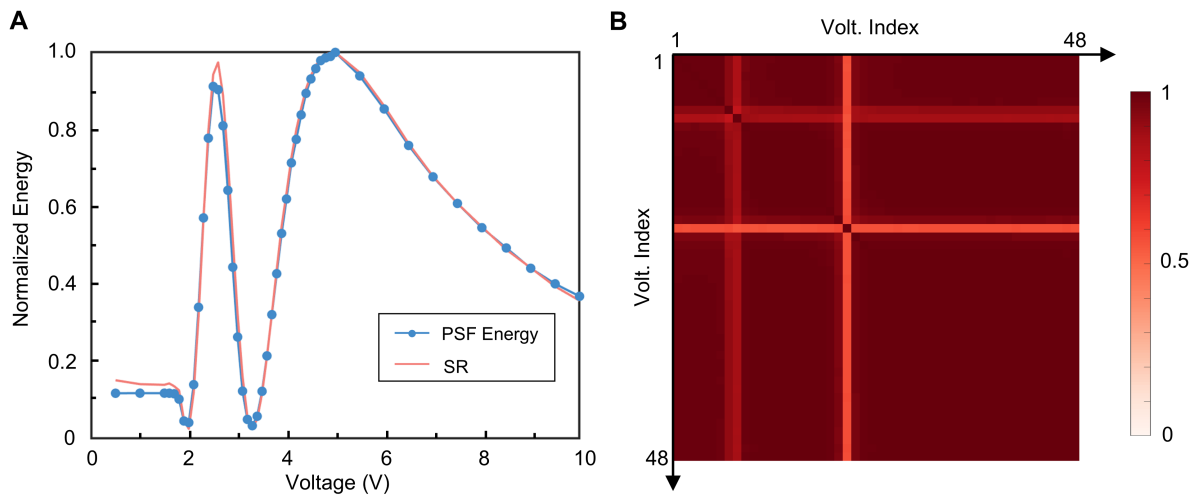

**Fig. S8. Analysis of PSF images' variation versus applied voltage.** (**A**) Intensity variation of the PSF images. The red line shows the calibrated spectral response data at 630 nm, and the blue asterisk line is plotted with the sum of all pixels' greyscales in the PSF image. (**B**) Correlation coefficient map of the captured PSF images. The voltage index (1 to 48) corresponds to the voltage sequence {0.5V, 1.0V, …, 10V} in Fig. S7.

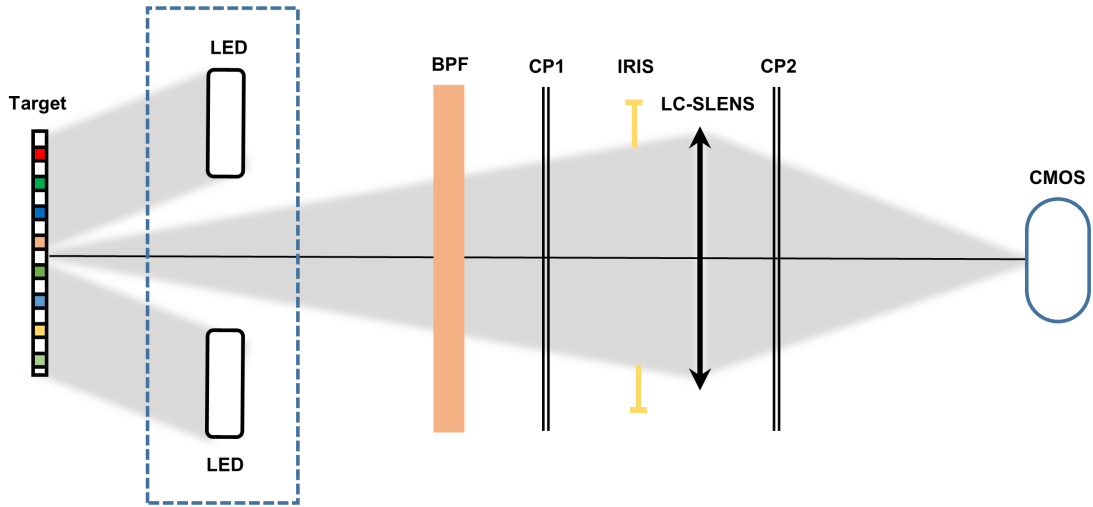

**Fig. S9. Optical setup for spectral imaging with LC-SLENS.** For targets like poster, LED light sources are used for illumination, while external illumination is not required in the experiment with micro-LED screen.

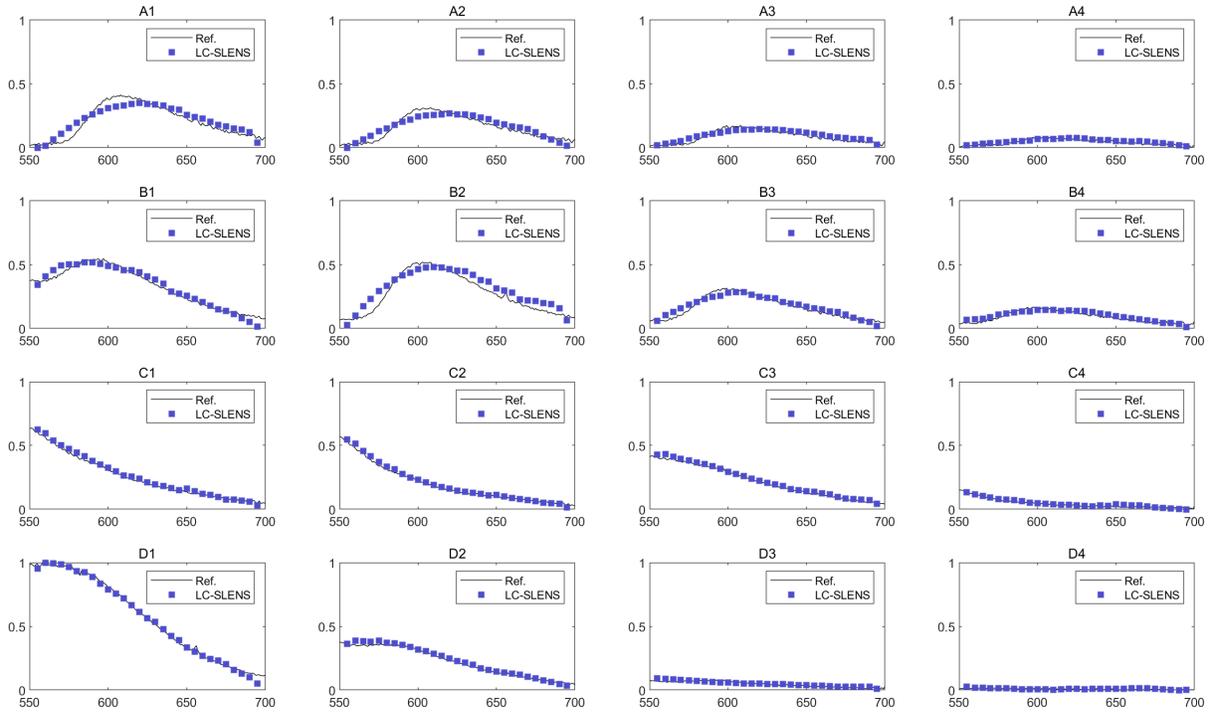

**Fig. S10. Spectral profiles of 16 color blocks on the color board.** The spectra measured by a commercial spectrometer (Ref.) and the reconstructed spectra with our LC-SLENS are denoted by black line and blue square, respectively. The reconstructed spectra for all color blocks agree well with the results obtained using a commercial spectrometer.

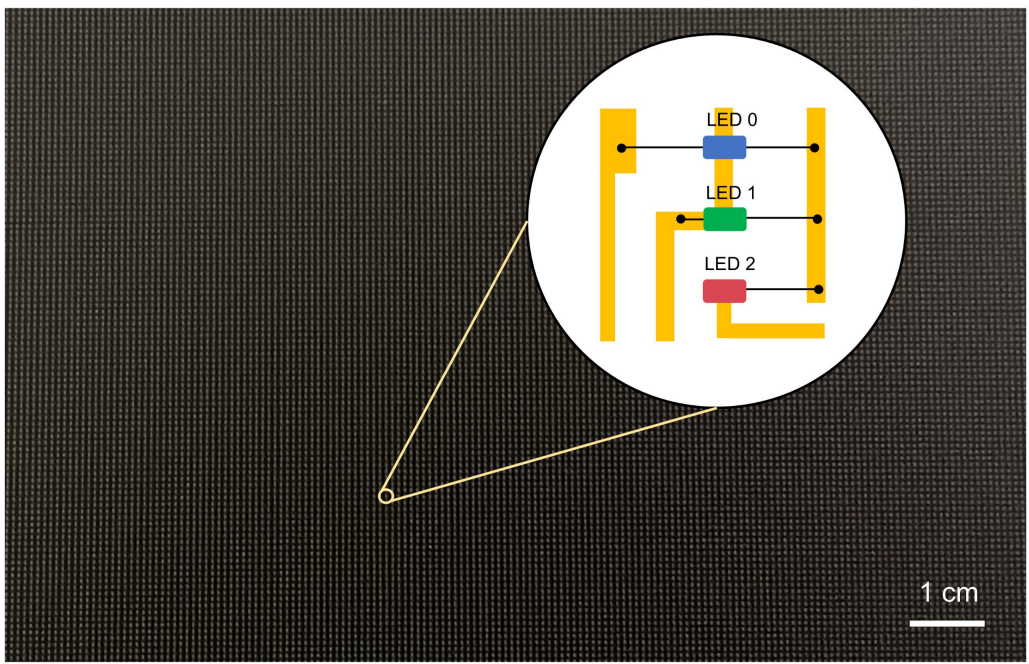

**Fig. S11. Optical image of the micro-LED screen.** The pixel pitch of the screen is 0.78 mm. Chip on board (COB) package is used for the blue, green and red LEDs in each pixel and the configuration is schematically denoted in the zoom-in circular area.

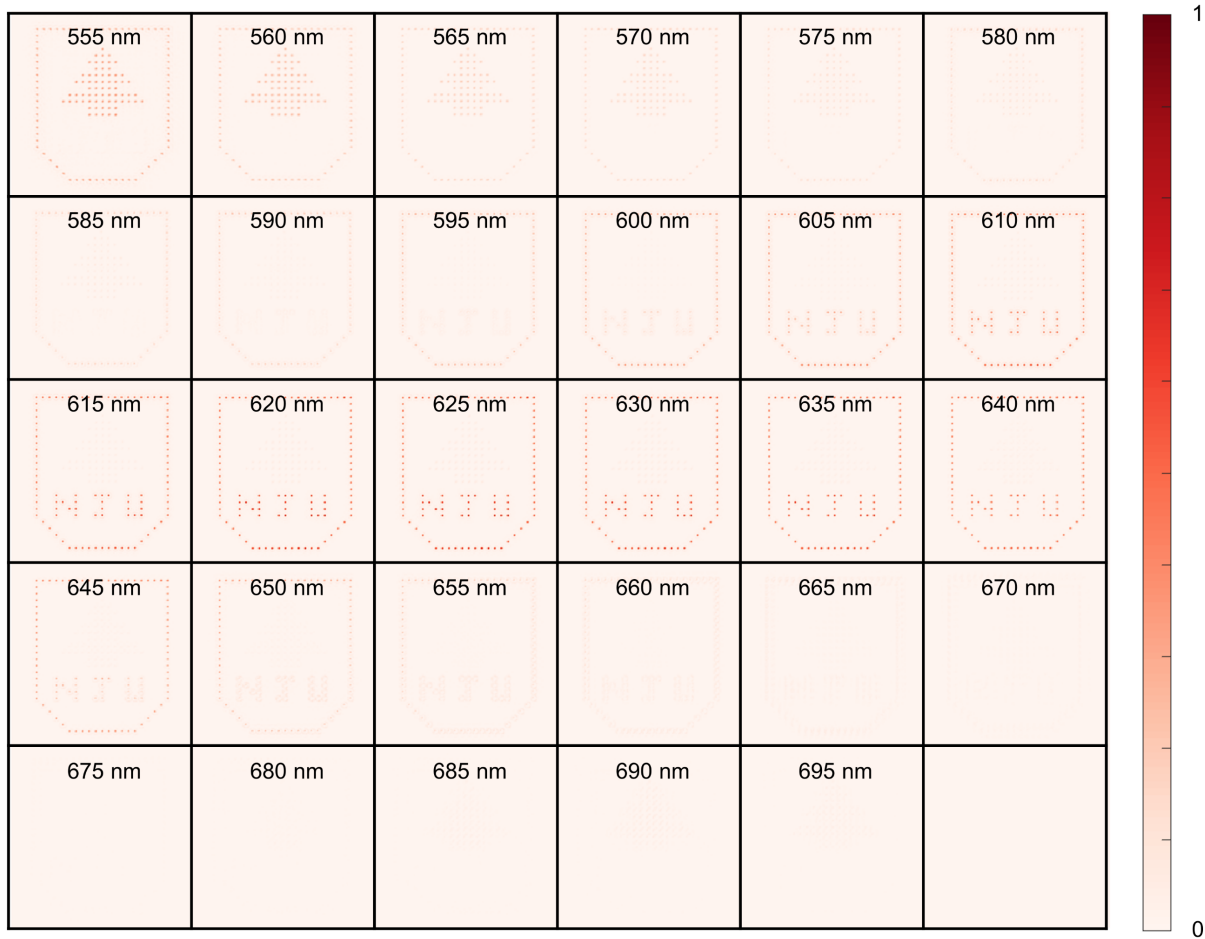

**Fig. S12. Spectral images of the screen pattern obtained with our hyperspectral camera.** The spectral images are normalized according to the maximum intensity in the spectral datacube.

**Table S1. Quantitative comparison of different schemes for miniaturized spectral imaging**

| Ref. | Working principle | | Wl. Range (nm) | Spec. Reso. (nm) | Spat. Reso. (μm) | Spectral images Pixel No. | Footprint (μm²) |
| --- | --- | --- | --- | --- | --- | --- | --- |
| | Spectral detection | Imaging | | | | | |
| *Sci. Adv.* **6**, eabc7646 (2020) | FP cavity | Metalens array | 795-980 | 9.4 | NA | NA | 2000×2000 |
| *ACM Trans. Graph.* **38**, 1–13 (2019). | DOE lens with spectrally-varying PSFs | | 420-660 | 10 | NA | 1440*960 | 1000×1000 |
| *Nat. Commun.* **13**, 2732 (2022). | Transversely dispersive metalens array | | 450-650 | 4 | 2.9 | 75*75 | 1440×1440 |
| *Nature.* **523**, 67–70 (2015). | Quantum dot array | NA | 390-690 | 3.2 | NA | NA | ~$10^4 \times 10^4$ [d] |
| *Nat. Commun.* **10**, 1020 (2019). | Photonic crystal slab | Supercell Repetition | 550-750 | 1 | 210[a] | 10*10 | ~2000×2000 |
| *Science.* **365**, 1017–1020 (2019). | Nanowire | Spatial Scanning | 500-630 | 10 | 50-100[a] | ~30*30 | 0.5×75[d] |
| *Nat. Photonics.* **15**, 601–607 (2021). | BP heterostructure + Bias volt. | NA | 2000-9000 | 420 | NA | NA | 9×16[d] |
| *Science.* **360**, 1105–1109 (2018). | BIC Metasurface | NA | 5700-7400 | ~21 | NA | NA | 1500×1500[d] |
| *Optica.* **9**, 461 (2022). | Reconfigurable Metasurface | Supercell Repetition | 450-750 | 0.8 | 87.9[a] | 356*436 | <7000×7000 |
| *eLight.* **2**, 23 (2022). | LC Metasurface + Bias volt. | NA | 1420-1470 | 9 | NA | NA | NA |
| *Adv. Mater.* **34**, 2200221 (2022). | Perovskite + Bias volt. | Spatial Scanning | 350-750 | 5 | 440[a] | ~5*5 | 440×440[d] |
| *Nat. Commun.* **13**, 4627 (2022). | vdW heterostructure + Bias volt. | Spatial Scanning | 1150-1470 | 20 | ~10[a] | NA | <10×10[d] |
| *Science.* **378**, 296–299 (2022). | vdW junctions + Bias volt. | Spatial Scanning | 405-845 | 3 | ~10[a] | 10*12 | 8×22[d] |
| *Nat. Photonics.* **17**, 218–223 (2023). | FP cavity | Supercell Repetition | 450-650 | 10 | 339 | 1920*1080 | 2560×2560 |
| **This work** | **LC spectral lens + Bias Volt.** | | **550-700** | **5** | **250/31[b]** | **500*500[c]** | **2120×2120** |

(a) the highest spatial resolution determined by the principle

(b) X/Y: resolution in object space (X) and image space (Y)

(c) determined by FOV

(d) footprint in spectrometer configuration (one-point spectral acquisition)

**Movie S1: Captured image frames of a color board when different voltages are applied on the LC-SLENS.** The corresponding voltage is denoted in the upper right corner (from 5V to 2V and with a gap of 0.1V).

**Movie S2: Captured image frames of a USAF 1951 resolution chart when different voltages are applied on the LC-SLENS.** The corresponding voltage is denoted in the upper right corner (from 5V to 2V and with a gap of 0.1V).